\documentclass{aa}
\usepackage{graphicx}
\usepackage{natbib}
\bibpunct{(}{)}{;}{a}{}{,}

\def\Msun{$M_\odot$}
\def\Neff{$N_\mathrm{eff}$}
\def\Ntot{$N_\mathrm{tot}$}
\def\lacute{\mathopen{<}}
\def\racute{\mathclose{>}}
\def\lbig{{\mathrm L}_{\lambda}}
\def\lcal{{\cal L}_{\lambda}}
\def\lsmalli{\ell_{\lambda \, i}}
\def\lsmallj{\ell_{\lambda \, j}}

\begin{document}

\title{Confidence levels of evolutionary synthesis models}
\subtitle{II: On sampling and Poissonian fluctuations}

\author{M. Cervi\~no\inst{1,2,3}
\and D. Valls-Gabaud\inst{1}
\and V. Luridiana\inst{4,5}
\and J.M. Mas-Hesse\inst{6}
}
\institute{UMR CNRS 5572, Observatoire Midi-Pyr\'en\'ees, 14, avenue Edouard Belin,
               31400 Toulouse, France \and
           Centre d'Etudes Spatiales des Rayonnements, CNRS/UPS, B.P.~4346,
               31028 Toulouse Cedex 4, France \and
           Max-Planck-Institut f\"ur extraterrestrische Physik,
              Giessenbachstrasse, 85748 Garching, Germany \and
           Instituto de Astronom\'\i a, UNAM, Apdo. Postal 70-264,
               04510 M\'exico D.F., Mexico \and
           European Southern Observatory, Karl-Schwarzschild-Str. 2,
              D-85748 Garching bei M\"unchen, Germany \and
           Laboratorio de Astrof\'\i sica Espacial y F\'\i sica Fundamental
           (LAEFF-INTA), POB 50727, E-28080 Madrid, Spain
}

\offprints{Miguel Cervi\~no}
\mail{mcs@laeff.esa.es}
\date{Received ; accepted }
\authorrunning{M.~Cervi\~no et al.}
\titlerunning{Confidence levels in population synthesis II}

\abstract { In terms of statistical fluctuations, stellar population
synthesis models are only asymptotically correct in the limit of a large
number of stars, where sampling errors become asymptotically small. When
dealing with stellar clusters, starbursts, dwarf galaxies or stellar
populations within pixels, sampling errors introduce a large dispersion in
the predicted integrated properties of these populations.  We present here
an approximate but generic statistical formalism which allows a very good
estimation of the uncertainties and confidence levels in any integrated
property, bypassing extensive Monte Carlo simulations, and including the
effects of partial correlations between different observables. Tests of the
formalism are presented and compared with proper estimates.  We derive the
minimum mass of stellar populations which is required to reach a given
confidence limit for a given integrated property.  As an example of this
general formalism, which can be included in any synthesis code, we apply it
to the case of young ($t \leq 20 $ Myr) starburst populations.  We show
that, in general, the UV continuum is more reliable than other continuum
bands for the comparison of models with observed data. We also show that
clusters where more than 10$^5$ M$_\odot$ have been transformed into stars
have a relative dispersion of about 10\% in Q(He$^+$) for ages smaller than
3 Myr. During the WR phase the dispersion increases to about 25\% for such
massive clusters.  We further find that the most reliable observable for
the determination of the WR population is the ratio of the luminosity of
the WR bump over the H$\beta$ luminosity.  A fraction of the observed
scatter in the integrated properties of clusters and starbursts can be
accounted for by sampling fluctuations.   }

\maketitle

\section{Introduction and motivation}

In the past few years, the increasingly detailed observations of stellar
populations in a wide variety of environments -- from stellar clusters to
high redshift galaxies -- has driven the development of increasingly
complicated evolutionary synthesis codes. A great care has been given to
the use of updated physical ingredients (like more realistic model
atmospheres, grids of observed spectra, stellar tracks covering more
evolutionary phases, increased wavelength range, etc) and technical details
(interpolation methods, convolutions, etc), leading to a situation where
the outputs produced by different codes are roughly consistent with each
other, although they perhaps disagree on the interpretation of a given
observable.  Systematic internal errors can be addressed comparing the
outputs resulting from different inputs or assumptions in a given code
\citep[e.g.,][]{Bru2001}, while the cross-comparison of different codes
with the same inputs gives an estimate of the external errors
\citep[e.g.,][]{Leith96}.

In contrast, the study of statistical errors has received comparatively
little attention, even though it seems likely that the observed scatter in
the observed properties of systems with a relatively small number of stars
(stellar clusters, starbursts, dwarf galaxies or stellar populations in
pixels), can be accounted for by sampling fluctuations from a given (and
perhaps universal) stellar initial mass function (IMF).

In this context, \citet{BB77} pioneered the estimation of the dispersion of
observables in stellar clusters by computing Monte Carlo simulations
constrained to contain a given number of post- main sequence (PMS) stars
(so as to reproduce the number of observed PMS stars in the
colour-magnitude diagram of a given cluster). They concluded that under the
assumption of a Salpeter IMF, the observed scatter in the integrated
colours of young open clusters could be reproduced by their models. Updated
analyses, by \citet{CBB88}, \citet{GiBi93} and \citet{Bro99} among others,
yield the same results: the dispersion in a colour-colour or
colour-magnitude diagram can be accounted for by these stochastic effects,
at least in part.

At older ages, the presence of a few cool but very luminous stars produces
the same effects. In the near-infrared, the small number of thermally
pulsating AGB stars can produce strong fluctuations in the JHK bands, as
pointed out by \citet{SF97} \citep[see also ][ for a dissenting
view]{LM99}.  Even in the optical bands, the integrated colours of globular
clusters can be strongly influenced by the presence of post AGB stars, as
\citet{Bro99} have shown.

A similar problem arises when dealing with stellar populations in pixels,
particularly in images or frames of nearby galaxies where each pixel could
include a limited number of stars which however account for most of a given
observable.  See \citet{Buzz93} and \citet{Ren98} for further details.

In the context of starbursts, \citet{May95} performed a similar analysis,
this time constraining the number of main sequence stars with spectral type
earlier than O4, finding that while the errors in colours were larger when
red supergiants appeared, they decreased when the mass of the cluster
increased, being negligible for clusters more massive than about 10$^5$
\Msun.

There is, however, no systematic study of the effects of these sampling
fluctuations on the integrated properties (such as colours or equivalent
widths of emission and absorption lines) of simple stellar
populations. \citet{Buzz89} pioneered this field by presenting a simple
formalism based on Poissonian errors, and applied it to synthesis models of
old stellar populations. Although powerful, this formalism does not account
for some important sources of errors and is not entirely rigorous.

While it is clear that the foreseen improvements in synthesis codes will
attempt to refine their realism, the detailed comparison with observed
properties will still suffer from the limited number of stars which
contribute to some observables.  The present paper is the second of a
series whose objective is to study the accuracy of evolutionary synthesis
models when compared with observational data, and to point out the
oversimplifications used in such models. This on-going project reviews
current problems in synthesis codes and shows possible ways to overcome
them, providing a guideline to include improvements in any synthesis code.

The first paper \citep{CLC00} pointed out the problem of confidence limits
in synthesis models set by the presence of statistical fluctuations in the
IMF, where the fluctuations were estimated by Monte Carlo simulations.  In
this work we present a statistical formalism to estimate {\it
quantitatively} the fluctuations expected in some of the most relevant
observables, and apply it, as an illustrative example, to the case of young
stellar populations (ages smaller than 20 Myr), produced by statistical
fluctuations in the IMF.  The effect of Poissonian statistics on
time-integrated observables such as supernovae rates, integrated mechanical
energy output and chemical yields, as well as major problems in the time
interpolations used in synthesis codes, are addressed in a third paper
\citep{Cetal01b}. Improvements in track interpolation techniques will be
discussed in a forthcoming paper.  We will complete the series by extending
the results presented in this paper to different star formation scenarios.

The structure of the paper is the following. In Sect. \ref{sec:pois} we
present our statistical formalism, which can be implemented in any
synthesis code. In Sect. \ref{sec:model} we use the formalism to derive the
dispersion of some relevant observables of young star-forming regions,
highlighting the importance of a proper use of covariance factors in the
evaluation of the dispersion. In Sect.  \ref{sec:Montecarlo} we compare the
analytical dispersion and confidence levels obtained with our formalism
with the ones obtained via Monte Carlo simulations.  We discuss the results
and draw our conclusions in Sect. \ref{sec:dis}. In Appendix \ref{ap:err}
we detail error propagation in observables, giving practical examples.  All
the results of the paper are available from our web server at {\tt
http://www.laeff.esa.es/users/mcs}.

\section{On sampling and Poissonian fluctuations}
\label{sec:pois}

Let us assume that \Ntot ~stars are observed, with masses distributed
between $m_\mathrm{low}$ and $m_\mathrm{up}$. The mass $m_i$ of the $i$-th
star is a random variable whose probability distribution function is given
by $\Phi(m_i)$ the stellar initial mass function. That is,

\begin{equation}
\int_{m_\mathrm{low}}^{m_\mathrm{up}} \; \Phi(m) \; dm \; \equiv  \; 1
\label{eq:pdfimf}
\end{equation}

For the case of a single power law IMF in that mass range, $\Phi(m) \sim
m^{-\alpha}$, so that masses can be generated by a simple transformation
from a random variable $u$ uniformly distributed in the interval $[0,1]$
with

\begin{equation}
m \; = \; \left[ (1-u) \, m_\mathrm{low}^{1-\alpha} \; + \;
      u \, m_\mathrm{up}^{1-\alpha}\right]^{1/(1-\alpha)}
\label{eq:imfsingle}
\end{equation}

\noindent and similarly for piece-wise power laws in different mass ranges.
Codes which use Monte Carlo sampling of the IMF must follow the evolution
of each star generated in this way.

The number $N_i$ of stars of mass $m_i$ is another random variable, whose
distribution function follows Poisson statistics. That is, the only
parameter of the Poisson distribution is precisely the value of the IMF at
that mass

\begin{equation}
d n_i \, = \, d N_i \, / \, N_\mathrm{tot} \; = \; \Phi(m_i) \, d m_i
\label{eq:pdfn}
\end{equation}

\noindent since obviously the total number of stars is $\int d N_i \equiv $
\Ntot .  Note that stellar evolution does not change the Poissonian nature
of $N_i$, and that the normalised random variable $n_i$ is almost a Poisson
variable, since the ratio of a Poisson variable with a
constant\footnote{Formally \Ntot ~is also a random variable (the sum of
random variables) but we assume here that is is fixed, that is, there are
actually only \Ntot -- 1 independent variables. For \Ntot $\sim 10^3$ or
more this makes a tiny difference and greatly simplifies the formalism.} is
not necessarily a Poisson variable \citep[e.g.,][]{KS77}.

\begin{figure}
\centering \includegraphics[angle=270,width=7cm]{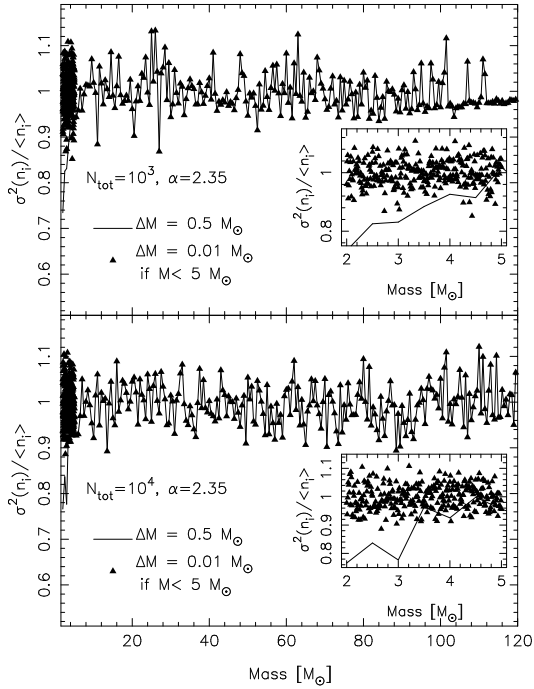}
\caption{Ratio of variance over mean value as a function of mass for 1000
Monte Carlo simulations with \Ntot~ = $10^3$ stars (upper panel) and \Ntot~
= $10^4$ stars (lower panel). Solid line: $\sigma^2(n_i)/n_i$ using a
mass-bin of 0.5 M$_\odot$ for the entire mass range. Triangles:
$\sigma^2(n_i)/n_i$ using a mass bin of 0.5 M$_\odot$ in the mass range
$[5,120]$ M$_\odot$, and a mass bin of 0.01 M$_\odot$ when 
$2 \mathrm{M}_\odot< m < 5 \mathrm{M}_\odot$.}
\label{fig:sig3} 
\end{figure}

A simple test to check the Poissonian nature of the distribution of $n_i$
is simply the ratio of its variance $\sigma^2(n_i)$ to its average value
$\lacute n_i \racute$ as a function of mass $m_i$. This ratio should be
close to unity for the entire range of masses considered. The situation is
illustrated in Fig. \ref{fig:sig3} with 1000 Monte Carlo simulations of
clusters with $10^3$ and $10^4$ stars with a Salpeter IMF slope
($\alpha=2.35$) in the mass range 2 -- 120 M$_\odot$.  Figure
\ref{fig:sig3} shows that despite the fact that $n_i$ is the ratio of a
Poisson variable with a constant, it is Poisson-distributed to within 10\%
even in clusters with as few as \Ntot = 10$^3$ stars.
Note that codes which use a binning in mass must ensure that the size
of the bins is small, otherwise the correlations induced may produce
a systematic effect, especially for the lower mass stars, as shown
on Fig.~\ref{fig:sig3}.

The fact that $n_i$ can be approximated by a Poisson variable makes it
possible to apply a proper statistical formalism, although it is slightly
more subtle than one could naively expect. Take for instance the integrated
monochromatic luminosity $\lcal$ of the stellar population of \Ntot~ stars
at some wavelength $\lambda$ at some age $t$. This is simply given by the
linear combination of individual monochromatic luminosities $\lsmalli$ of
stars of mass $m_i$ at the evolutionary stage of age $t$:

\begin{equation}
\lcal(t) \; = \; N_\mathrm{tot} \, \sum_{i=1}^{N_\mathrm{tot}}
\, n_i \, \lsmalli(t)
\label{eq:lum}
\end{equation}

For a given age $t$, $\lsmalli$ is a constant, so that $\lcal$ is a
weighted sum of Poisson-distributed random variables, and would itself be a
Poisson variable if the weights would take integer values, since the sum of
Poisson variables is also a Poisson variable. Hence define
$\mathrm{B}_{\lambda}$ as

\begin{equation}
\mathrm{B}_{\lambda} \, = \, 1 \, + \mathrm{int} \left[ \mathrm{min}\left(
\lsmalli \right) \right]
\end{equation}

\noindent so that

\begin{equation}
\lcal(t) \; = \; N_\mathrm{tot} \, \left( 1 + \mathrm{B}_{\lambda} \right)
\; \sum_{i=1}^{N_\mathrm{tot}} \, n_i \, \frac{ 1 +
\mathrm{int}\left(\lsmalli(t)\right)}{1 + \mathrm{B}_{\lambda}}
\end{equation}

In this case, and up to a constant, $\lcal$ would be a proper Poisson
variable, the (integer) sum of Poisson variables, with parameter given by
the sum of the individual Poisson parameters. However, as long as we are
interested in {\it dispersions} and not actual distribution functions, we
can disregard the actual distribution function of $\lcal$ as long as it is
a {\it linear} combination of random variables.  It is convenient to
introduce at this stage the random variable

\begin{equation}
w_i \; \equiv \;
   n_i \, N_\mathrm{tot} \, / M_\mathrm{tot} \; = \; N_i \, / \, M_\mathrm{tot}
\label{eq:defwi}
\end{equation}

\noindent which gives the normalised mean value of the number of stars of
mass $m_i$ to the total mass of the cluster $M_\mathrm{tot} = \sum m_i$
assumed to be a constant\footnote{Again, formally $M_\mathrm{tot}$ is also
a random variable with a mean value, for a Salpeter IMF between 2 and 120
M$_\odot$, of $\lacute M_{\mathrm tot} \racute = 5.97 N_\mathrm{tot}$ and a
variance ${\mathrm{var}}(M_{\mathrm tot})= 76.28 N_\mathrm{tot}$, but we
will assume here that it is a constant to simplify the formalism.}. The
integrated monochromatic luminosity per unit stellar mass, or specific
luminosity, is

\begin{equation}
\lbig \; = \; \lcal \; / \; M_\mathrm{tot}
\end{equation}

With this notation, the expectation value of $\lbig$ at some age $t$
becomes

\begin{equation}
\lacute {\mathrm L}_{\lambda} \racute \; = \; \sum_i \, w_i \, \lsmalli
\label{eq:avelum}
\end{equation}

The actual integrated specific luminosity $\lbig$ will fluctuate around
this average value with a variance given by

\begin{eqnarray}
\mathrm{var}({\mathrm L}_{\lambda})  & =&
     \; \lacute \left( {\mathrm L}_{\lambda} -
     \lacute {\mathrm L}_{\lambda} \racute \right)^2 \racute \; \nonumber \\
    & = & \; \sum_i \, w_i \, \lsmalli^2
     \; + \; \sum_{i\not=j} \, \lsmalli \, \lsmallj \, \mathrm{cov}(w_i,w_j)
\label{eq:varlum}
\end{eqnarray}

\noindent where the covariance is given, as usual, by

\begin{equation}
\mathrm{cov}(w_i,w_j) \; \equiv \; \lacute \left( w_i - \lacute w_i \racute
\right) \, \left( w_j - \lacute w_j \racute \right) \racute \;
\label{eq:defcovw}
\end{equation}

Note that Eq. \ref{eq:varlum} is an {\it exact} result since $\lbig$ is a
{\it linear} function of $w_i$. The random variables $w_i$ and $w_j$ are
independent, so their covariance vanishes. Note however that codes which
produce bins of mass and follow the average evolution in that bin introduce
a non-zero covariance which must be taken into account, particularly if the
bin size is large. We have performed several tests with different mass
binning, and found that when the mass bin is smaller than 0.5
M$_\odot$, the covariance terms can be neglected for M$>$5 M$_\odot$.
Furthermore, if the mass bin is made less or equal to 0.1 M$_\odot$, the
covariance terms also vanish in the mass range 2--5 M$_\odot$.

So far, we have shown that $\lbig$ (derived from $\lcal$) is almost a
Poisson variable, so we may expect its variance $\sigma^2(\lbig)$ to be
similar to its mean. This will not be exactly so, as discussed above.
There are two possibilities here. The first one is to assume that the
Poisson distribution is a good approximation to the distribution of
$\lbig$. In this case, we can derive approximate confidence limits in the
standard way \citep[e.g.,][]{Geh86} and then infer confidence limits for
derived quantities, such as magnitudes, colours, equivalent widths,
etc. This proves to be rather cumbersome. The second possibility is to
disregard the actual distribution function of the specific integrated
monochromatic luminosity, and deal with variances only, which can then be
propagated for any derived quantity (see Appendix A).

In this context we can introduce, following \citet{Buzz89}, the concept of
an {\it effective number} of stars $N_\mathrm{eff}^*({\mathrm L}_\lambda)$,
from the relative error in the integrated monochromatic luminosity

\begin{eqnarray}
\frac{1}{\sqrt{N_\mathrm{eff}^*(\mathrm{L}_\lambda)}} & = &
\frac{{\mathrm{var}}({\lbig})^{1/2}}{\lbig} = \frac{\left(\sum_i
\mathrm{var}(w_i) \lsmalli^2\right)^{1/2}}{ \sum_i w_i \lsmalli } =
\nonumber \\ & = & \frac{1}{\sqrt{M_\mathrm{tot}}} \frac{(\sum_i w_i
\lsmalli^2)^{1/2}}{\sum_i w_i \lsmalli}
\label{eq:neff}
\end{eqnarray}

\noindent where we have made use of the fact that ${\mathrm{var}}(w_i) =
\mathrm{var}(N_i) / M_\mathrm{tot}^2 = w_i / M_\mathrm{tot} $ given that it
is approximately Poisson. Even if the distribution function is not Poisson,
one can {\it still} define this effective number as an approximate Poisson
dispersion.

As \citet{Buzz89} emphasized, $N_\mathrm{eff}^*$ does not correspond to a
real number of stars, but rather to an estimation of the relative error, in
the Poisson limit. It can be interpreted as a number of contributors to a
given wavelength at that age, so that if this number is small, we may
expect large fluctuations in the mean value of the luminosity at this
wavelength.  In the case where all the stars have the same luminosity,
$\lsmalli = l_\lambda$, and since $\sum w_i =$ \Ntot~~/$M_\mathrm{tot}$
then $N_\mathrm{eff}^*(\lbig) = N_\mathrm{tot}$, and all stars contribute
the same, obviously.  In the general case, however,
$N_\mathrm{eff}^*(\lbig)$ can never exceed \Ntot, by definition. Hence
small values indicate that the average value must be interpreted with
caution, the variance around the mean value being as large as the mean
value, typically.  The ratio $N_\mathrm{eff}^*({\mathrm
L}_{\lambda})/N_\mathrm{tot}$ can also be interpreted as a measure of the
statistical entropy of the $\lsmalli$ values \citep{Buzz93}: for a given
$M_\mathrm{tot}$, a larger value of $N_\mathrm{eff}^*({\mathrm
L}_{\lambda})$ means a lower dispersion in $\lsmalli$.  We can also rescale
$N_\mathrm{eff}^*$ to the total mass in stars, following the convention
used in population synthesis :

\begin{equation}
N_\mathrm{eff}=N_\mathrm{eff}^* \;  / \; M_\mathrm{tot}
\end{equation}

\noindent and the definition of $N_\mathrm{eff}^*$ can be generalized to
almost all observables of a synthetic population (ratios, equivalent
widths, colours etc., see Appendix A for more details).  This formulation
has two main advantages:

\begin{itemize}
\item For a cluster of a given mass, $N_\mathrm{eff}$ gives a simple and
easy to estimate measure of the expected dispersion of a given observable.
This dispersion depends on the number of stars in the cluster which
effectively contributes to that observable. It must always be computed if a
meaningful comparison with observed quantities of clusters of that mass is
made.
\item The comparison of the different values of $N_\mathrm{eff}$ for
different observables (say, luminosities in two different passbands) gives
a direct estimation of the difference in the statistical entropy of the
observables, which helps to establish which observable is more reliable and
thus better constrained by observations.
\end{itemize}

The $N_\mathrm{eff}$ formalism can, and should, be included in any
synthesis code, and the corresponding values of
$N_\mathrm{eff}(\mathbf{x})$ must be given along with every computed
observable $\mathbf{x}$ so that the dispersion in $\mathbf{x}$ can be
assessed.  Before addressing the comparison of the dispersions computed
with this analytical formalism with the ones obtained more precisely via
Monte Carlo simulations in Sect. \ref{sec:Montecarlo}, it is worth giving a
practical example of its application.

\section{A worked example : the case of young starbursts}
\label{sec:model}

The first application of a simplified version of this formalism was made by
\citet{Buzz89,Buzz93} in his study of old stellar populations, where the
presence --or otherwise-- of some evolutionary phases affected important
observables in the optical (see also \citet{LM99} for the case of infrared
colours). Rapid evolutionary phases also occur in young starburst
populations, and so as an illustration of the usefulness of the formalism,
we apply it here to an updated version of the code presented in
\citet{MHK91,CMH94,CMHK01}. The main updates to the code include the
following :

\begin{enumerate}
\item Inclusion of the full set of Geneva evolutionary tracks including
standard and enhanced mass-loss rates \citep{Meyetal94}.
\item Inclusion of metallicity-dependent atmosphere models for normal stars
from Kurucz and CoStar \citep{SK97} and the atmosphere models for WR stars
from \citet{Schmetal92}.
\item Inclusion of an analytical IMF formulation using a modified dynamical
mass-bin from the Dec. 2000 release of {\it Starburts99} \citep{SB99}. We
have also kept the original Monte Carlo formulation.  The dynamical
mass-bin subroutine used here makes use of a more restrictive mass binning
than the one used in {\it Starburts99} with a higher numerical precision in
order to ensure a narrow enough mass bin for the computation of the
dispersion.  The typical size of the dynamic mass bin used in our code is
10$^4$ mass points. This numerical resolution does not produce however any
significant changes in the mean properties obtained by both codes.
\item Use of parabolic interpolations in time for the computation of the
isochrones (see \nocite{Cetal01b}Cervi\~no et al. 2001b, Paper {\sc iii} of
this series, for a full study and justification).
\end{enumerate}

The code also includes an intermediate track that takes into account the
discontinuity in the evolution of WR and non-WR stars. Such an intermediate
track was introduced by \cite{CMH94} and will be discussed in Paper {\sc
iv} of this series.

The complete predictions of the updated set of evolutionary synthesis
models will be presented elsewhere.  For the present example we will focus
on solar metallicity simple stellar populations with standard mass-loss
rates \citep{Schetal92}, a Salpeter IMF slope in a mass range 2 -- 120
M$_\odot$, an Instantaneous Burst star-formation law and ages from 0 to 20
Myr.  We will use throughout this work an analytical approximation of the
IMF, we will not use the intermediate tracks mentioned above, and we will
neglect the presence of binary systems, so that our dispersion estimates
can be compared to those obtained by other synthesis codes with similar
inputs.

In the following subsections we discuss the dispersion estimated for a few
observables predicted by the code: continuum luminosities, V--K colour,
H$\beta$ equivalent width and some WR ratios.  We focus here on the
evolution of the dispersion of these observables, but we also show their
general evolution in order to understand their dispersion. A detailed study
of the evolution of these observables has been discussed elsewhere (see
\citet{CMH94} for a more detailed description of the evolution, and
\citet{LH95}, for a comparison with other synthesis models, showing that
both codes give quite similar results)\footnote{The only relevant
difference between the solar metallicity models presented here and the ones
from the Dec. 2000 release of {\it Starburts99} ({\tt
http://www.stsci.edu/science/starburts99/ }) is the use of different
atmosphere models.  Incidentally, the results presented here can be used to
evaluate dispersions in the {\it Starburst99} predictions, when using the
same input parameters.}.

\subsection{Continuum properties}

\begin{figure*}
\centering \includegraphics[angle=270,width=17cm]{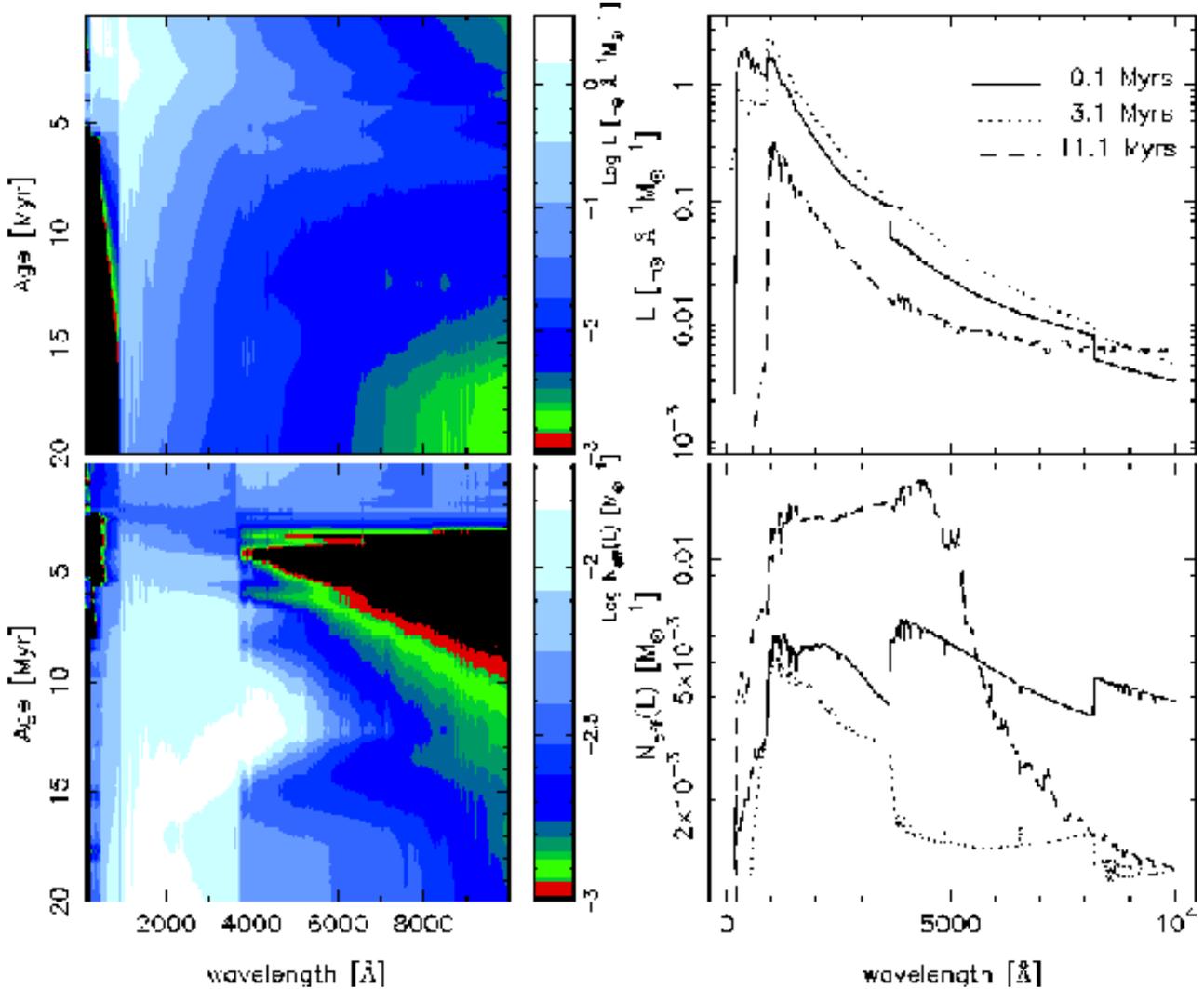}
 \caption{Specific continuum flux. {\it Top left:} Evolution of the overall
 spectrum.  Units are L$_\odot$ \AA$^{-1}$ M$_\odot^{-1}$.  {\it Bottom
 left:} Effective number of stars for the specific flux, in units of
 M$_\odot^{-1}$, for the overall spectrum. The right panels show the same
 quantities for selected ages.}  \label{fig:sed}
\end{figure*}

Equation \ref{eq:neff} can be applied in a straightforward way to the
continuum predicted by the synthesis code. We only need to take into
account that each $\lsmalli$ is the luminosity of a star of initial mass
$m_i$ with a relative weight given by $w_i$.

A small complication arises in the case of young star-forming regions in
that each massive star also contributes to the ionisation of the cloud
around the cluster, and thus to the nebular continuum. This contribution
depends on the properties of the gas in the nebula (the electronic
temperature $T_\mathrm{e}$, the electronic density $n_\mathrm{e}$, and the
metallicity, in the simplest case). We have used in the following the same
inputs as \citet{MHK91}: $T_\mathrm{e}=10^4$K, $n_\mathrm{e}=100\, {\mathrm
cm}^{-3}$ and {N(He {\sc ii})/N(H {\sc ii}) = 0.1}. We have also assumed
that only a fraction $f=$0.7 of the ionising photons is absorbed by the
gas.

In this case the luminosity $\lsmalli$ at a given wavelength has two
components, i.e. $\lsmalli = l_{*,i} + l_{n,i}$ where $l_{*,i}$ is the
stellar luminosity, and $l_{n,i}$ is the contribution of the star to the
nebular continuum. The scaled variance in the total (stellar plus nebular)
specific luminosity is then

\begin{eqnarray}
\sigma_{ \mathrm{L}_\mathrm{tot}}^2 & = &
     \mathrm{var}(\mathrm{L}_\mathrm{tot}) \, M_\mathrm{tot}^{-1} = \sum_i
     w_i \, \lsmalli^2 = \nonumber \\ &=&\sum_i w_i l_{*,i}^2 + \sum_i w_i
     l_{n,i}^2 + 2\,\sum_i w_i l_{*,i} l_{n,i}
\label{eq:lcov}
\end{eqnarray}

The last term of the equation is the covariance between stellar and nebular
continua for a given mass or star $i$ :

\begin{equation}
{\mathrm{cov}}({\mathrm{L}}_{*},{\mathrm{L}}_n) = \sum_i \, w_i \, l_{*,i}
\, l_{n,i}.
\end{equation}

Note also that the scaled dispersion of the specific monochromatic
luminosity $\sigma(\mathrm{L}_\mathrm{tot})$ has units of erg s$^{-1}$
\AA$^{-1}$ M$_\odot^{-1/2}$.

The overall evolution of the continuum from 100 \AA\ to 10$^4$ \AA\ is
illustrated in the upper left panel of Fig.~\ref{fig:sed}. At the beginning
of the burst, massive stars are the dominant source of luminosity from the
UV to the near infrared. The extreme-UV continuum (below 500 \AA) appears
with the first WR stars (at about 2.5 Myr) and lasts until the end of the
WR phase (at about 5 Myr). The lower left panel shows the evolution of
$N_\mathrm{eff}(\lbig)$ as a function of age and wavelength. The figure
shows that the extreme-UV continuum produced by WR stars has a very small
$N_\mathrm{eff}(\mathrm L_\lambda)$ value, reflecting small number
statistics and the statistical entropy of the luminosities of these stars
in the synthetic cluster at these wavelengths. It also shows a large
dispersion in the optical and near infrared fluxes at ages between 2.5 Myr
and 8.5 Myr.

The evolution of the continuum for some selected ages and the corresponding
$N_\mathrm{eff}(\mathrm L_\lambda)$ values are shown on the right panels of
Fig.~\ref{fig:sed}.  The comparison of the lower right panel with Fig. 2 of
\cite{Buzz89} is interesting.  Buzzoni presents $N_\mathrm{eff}(\mathrm
L_\lambda)$ for a 15 Gyr synthetic cluster, showing an abrupt change in
$N_\mathrm{eff}(\mathrm L_\lambda)$ at the Balmer jump.  In spite of the
differences in the synthesis codes, ages and metallicity, the same pattern
is found in our Fig. \ref{fig:sed}. The comparison of both figures reveals
however the different nature of the Balmer discontinuity for stellar
clusters with different effective temperatures.  In clusters dominated by
hot massive stars, the blue continuum is dominated by the nebular emission
and is more luminous than the red continuum, while clusters with cooler
stars show the opposite behaviour. This means that the discontinuity in
$N_\mathrm{eff}(\mathrm L_\lambda)$ at the Balmer jump reflects the
increase in the statistical entropy of the luminosities of the stars
(including the stellar and the nebular contributions) at this
wavelength. At small ages (0.1 Myr), the cluster is dominated by hot stars,
emitting shortwards of the Balmer continuum: the statistical entropy for
$\lambda < \lambda ({\rm Bac})$ is lower than that found at $\lambda >
\lambda ({\rm Bac})$.  At 3.1 Myr, the nebular contribution to the
continuum decreases rapidly. The stellar continuum contributes most to the
total flux so that the statistical entropy for $\lambda < \lambda ({\rm
Bac})$ becomes now larger than at $\lambda > \lambda ({\rm Bac})$.  At 11.1
Myr, the remaining stars are so cool that the nebular emission no longer
plays a role in the Balmer discontinuity.  At older ages (cooler stars) the
Balmer discontinuity appears completely inverted. Let us illustrate this in
more detail by studying the correlation of the stellar and nebular fluxes :

\begin{itemize}
\item For a 0.1 Myr cluster, the nebular emission contributes more than
30\% to the total (stellar + nebular) flux at wavelengths longwards $\sim$
4000 \AA.  The emission at such wavelengths is therefore correlated with
the ionising flux of the stars which dominate the emission at shorter
wavelengths, and so $N_\mathrm{eff}(\mathrm L_\lambda)$ is essentially
flat.

\item At 3.1 Myr, the nebular emission contribution at 4000 \AA ~drops to
5\%.  The correlation with the ionising flux of massive stars disappears
and this is reflected by a decrease in the value of $N_\mathrm{eff}(\mathrm
L_\lambda)$.  Also, the emission of the different massive star populations
(i.e. the $\lsmalli$ values) is very variable and depends strongly on the
amount of stars that become WRs. Such a situation produces a large
dispersion in the ionising, optical and infrared continua.

\item The dispersion in the ionising continuum vanishes at ages older than
5 Myr (when the WR population disappears), but remains large in the optical
and infrared continua due to the presence of red supergiant (RSG) stars.
Note that WRs and RSGs have a rapid evolution, they are very luminous, and
a small amount of these stars may dominate the overall continuum. As long
as their total number is small, the intrinsic dispersion at such phases of
the evolution of the cluster will be large.

\item Finally, a 11.5 Myr old cluster shows a small Balmer jump, and all
the stars in the cluster have ``similar'' UV spectra,
i.e. $N_\mathrm{eff}(\mathrm L_\lambda)$ reaches a maximum value at this
wavelength. The spectrum of the individual stars becomes more and more
different towards the red, as shown by the decay of $N_\mathrm{eff}(\mathrm
L_\lambda)$ at longer wavelengths.
\end{itemize}

In general, the UV continuum has a larger $N_\mathrm{eff}(\mathrm
L_\lambda)$ value than the optical continuum (and thus a larger statistical
entropy). This also reflects the fact that the UV continuum is more
homogeneous along the spectral types, i.e. small differences in effective
temperature lead to similar UV spectra. The optical continuum depends more
on the effective temperature, and so small variations in the population
produce larger effects. Such a dependence becomes more and more important
towards longer wavelengths.

At the other extreme of the wavelength range, in the ionising continuum,
the situation is even more dramatic than at IR wavelengths. The differences
in the ionising continua of different stars are generally quite large, and
so this part of the continuum will be the most affected by statistical
fluctuations, as shown below.

\subsection{Ionising continuum }

\begin{figure}
 \resizebox{\hsize}{!}{\includegraphics[angle=270,width=7cm]{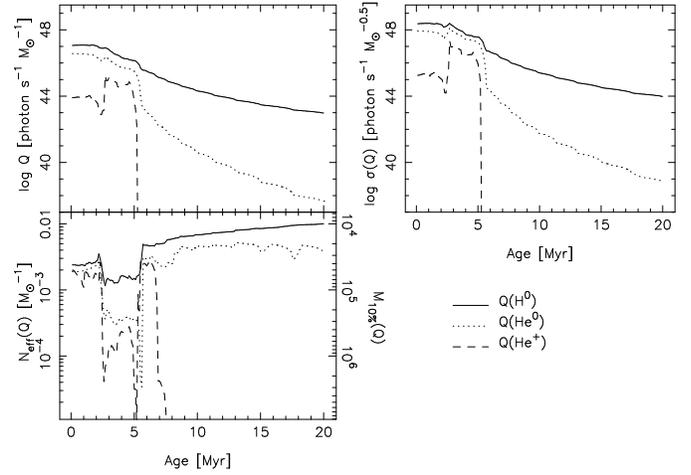}}
 \caption{Ionising photons. {\it Top left:} Evolution of the specific
 number of emitted ionising photons able to ionize He$^+$, Q(He$^+$),
 He$^0$, Q(He$^0$), and H$^0$, Q(H$^0$), in units of photons s$^{-1}$
 M$_\odot^{-1}$.  {\it Bottom left:} Effective number, in units of
 M$_\odot^{-1}$, for the different ionising photons (see text for
 definition). The right axis shows the minimum mass required to ensure a
 Poissonian dispersion smaller than 10\% assuming a mass range 2 -- 120
 M$_\odot$ with a Salpeter IMF slope.  {\it Top right:} $\sigma$(Q) in
 photons s$^{-1}$ M$_\odot^{-0.5}$ units for the same quantities.}
\label{fig:nlyc}
\end{figure}

We have calculated the predicted ionising flux shortwards of the He$^{+}$,
He$^{0}$ and H$^{0}$ edges, Q(He$^+$), Q(He$^0$), and Q(H$^0$)
respectively. The corresponding values as well as $N_\mathrm{eff}$(Q) and
$\sigma$(Q) are shown in Fig. \ref{fig:nlyc}. As before, the values shown
in Figure \ref{fig:nlyc} are normalised to the total mass transformed into
stars $M_\mathrm{tot}$.

 An important quantity in this context is the {\it minimum} mass of the
cluster required to ensure a dispersion smaller than 10\% in a given
observable, M$_{10\%}$. Clusters with smaller masses will have a dispersion
larger than 10\% in that observable. For instance, the right axis in the
bottom left plot of Fig.~\ref{fig:nlyc} gives the mass of the ensemble of
stars required to ensure a dispersion in the $Q$ values lower than 10\%,
denoted as M$_{10\%}$(Q). Note that the relation between the numbers of
stars present \Ntot\ and the total mass of the cluster depends on the IMF
and the mass limits adopted, and so the values of M$_{10\%}$ must be
rescaled if different assumptions are made regarding these values.

Note that the scaled dispersion $\sigma$(Q) is larger than the normalised
value of Q. It does not mean that we can obtain negative values of Q,
because the probability density distribution is quasi-Poisson, i.e. the
mean value and the corresponding error is not Q$\pm \sigma$(Q).  Take for
instance an extreme example, a cluster with a total mass
$\mathrm{M}_\mathrm{tot} = 120 \mathrm M_\odot$ made up by stars with
masses in the range 2--120 M$_\odot$ at an age of 0.1 Myr.  The predicted
value of the scaled (specific) average of Q(H$^0$) is 4$\times 10^{47}$
photons s$^{-1}$ M$_\odot^{-1}$, i.e. the actual value of Q(H$^0$) for such
a cluster would be 4.8$\times 10^{49}$ photons s$^{-1}$.  But the actual
Q(H$^0$) may vary between 0 photons s$^{-1}$ (if none of the stars in the
cluster produce the ionising flux) and $\sim 5 \times 10^{50}$ photons
s$^{-1}$ (if the cluster is formed by only one star of 120 M$_\odot$).  In
fact, the model predictions at the 90\% confidence level for such a cluster
is Q(H$^0$) = $4.8 \times 10^{49}\, ^{+4.1e50}_{-4.8e49}$ photons s$^{-1}$
(but see Sect. \ref{sec:Montecarlo} for details and cautions on such an
estimation).

This figure also shows another important feature: the relative dispersion
(the inverse of $N_\mathrm{eff}$) becomes larger for more energetic edges.
It is often assumed that the uncertainty on Q(H$^0$) is representative of
the uncertainty of the ionising continuum at shorter wavelengths. Figure
\ref{fig:nlyc} shows on the contrary that this assumption is false, since
the shorter the wavelength, the lower the related $N_\mathrm{eff}$(Q), and
the larger the relative dispersion.

Note that to ensure a relative dispersion of 10\% in the flux below the
Q(He$^{+}$) break more than 5 10$^4$ M$_\odot$ must be transformed into
stars (in the mass range 2 -- 120 M$_\odot$ with a Salpeter IMF).  The
situation becomes even worse when WR stars appear in the cluster. In this
case a mass of about 10$^6$ M$_\odot$ is needed to get an uncertainty of
about 10\%.  Note that the corresponding $N_\mathrm{eff}$(Q) values are
lower than the UV continuum ones, so that the dispersion in the emission
line spectrum is expected to be much larger than in the continuum.

In the past few years the continuum of evolutionary synthesis models has
been extensively used in the computation of grids of photoionisation
models.  Although the general predictions of photoionisation models can be
valid, the fitting of observed data with such grids, or the comparison of
the grids with a set of observed sources, must take into account the
possible dispersion due to these sampling fluctuations in the stellar
population.  The expected dispersion on the emission line spectrum will be
evaluated in a forthcoming paper.

\subsection{Monochromatic luminosities and colour indexes}

\begin{figure}
 \resizebox{\hsize}{!}{\includegraphics[angle=270,width=7cm]{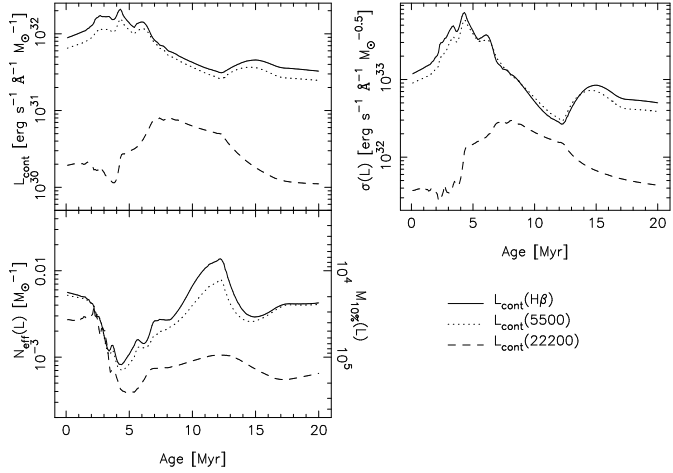}}
 \caption{Luminosities. {\it Top left:} Evolution of the specific
 monochromatic luminosity at 4861, 5500 and 22000 \AA ~in units of erg
 s$^{-1}$ \AA$^{-1}$ M$_\odot^{-1}$.  {\it Bottom left:} Effective number,
 in units of M$_\odot^{-1}$, for the different monochromatic
 luminosities. Right axis as in Fig.~\ref{fig:nlyc}. {\it Top right:}
 $\sigma(\mathrm L_\lambda)$ in units of erg s$^{-1}$ \AA$^{-1}$
 M$_\odot^{-0.5}$ for the same quantities.}  \label{fig:L}
\end{figure}

Figure \ref{fig:L} shows the evolution of the continuum luminosities at the
wavelength of the H$\beta$ line, at 5500 \AA ~(V band) and at 2.2 $\mu$m (K
band). The figure also shows $N_\mathrm{eff}(\mathrm L_\lambda)$ for the
given wavelengths and the required minimal cluster masses to ensure a
relative dispersion smaller than 10\%, M$_{10\%}(\mathrm L_\lambda)$, as
well as the associated dispersion $\sigma(\mathrm L_\lambda)$.

As it was pointed out above, at the beginning of the burst the nebular
contribution dominates, and so $N_\mathrm{eff}(\mathrm L_\lambda)$ is
similar at all wavelengths. Later on, $N_\mathrm{eff}(\mathrm L_\lambda)$
decreases when the fast evolving stars (WR and RSG) appear in the
cluster. The dispersion decreases for optical wavelengths when WR stars
disappear.

The results from this figure can be compared with Table 1 in \citet{LM99}.
Both results are quite in agreement, as well as with \citet{Buzz93}: IR
luminosities require in general a larger amount of mass transformed into
stars to ensure a given relative dispersion.  Equivalently, IR luminosities
are much more affected by the discreteness of the stellar population and
are more prone to sample fluctuations.

The minimal cluster masses required to ensure a dispersion smaller than
10\% for the 5500~\AA ~and 2.2~$\mu$m ~luminosities obtained by
\citet{LM99} are however a bit larger than the ones obtained by us. This
can be understood taking into account that they used a different mass range
(0.1 -- 120 M$_\odot$) and that they did not include the nebular
contribution in their calculations.

\begin{figure}
 \resizebox{\hsize}{!}{\includegraphics[angle=270,width=7cm]{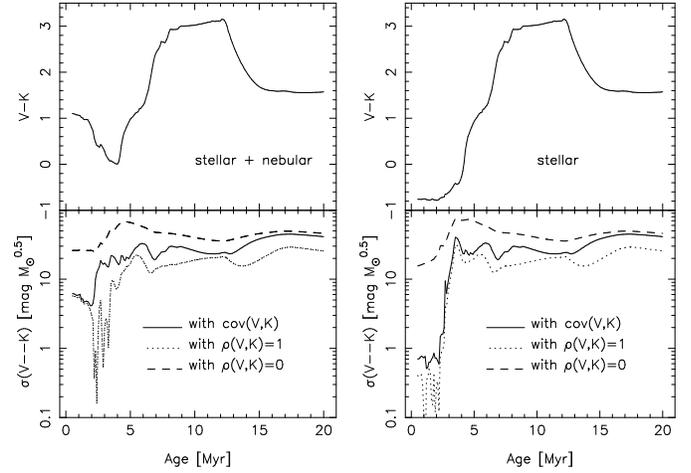}}
 \caption{V--K colour and its covariance. {\it Top panel:} Evolution of the
 V--K colour index for a cluster where the nebular contribution is included
 (left) and for a cluster with the stellar contribution only (right). {\it
 Bottom panels:} $\sigma$(V--K) in units of mag M$_\odot^{0.5}$ obtained
 with different assumptions about the covariance terms for the two
 scenarios mentioned above.}
\label{fig:app1}
\end{figure}

In Fig. \ref{fig:app1} we show the evolution of the V--K colour index and
its (scaled) dispersion. Since the results depend strongly on the proper
evaluation of the covariance factors, we have also considered the resulting
V--K colour index for the case where the stellar contribution is the only
one used.

Note that the normalisation units of $\sigma$(V--K) are mag M$_\odot^{0.5}$
rather than mag M$_\odot^{-0.5}$. As shown in Appendix A, this is the case
for the dispersion in {\it ratios} : the larger the mass, the lower the
dispersion.  The normalised $\sigma$(V--K) value at 20 Myr is about 40 mag
M$_\odot^{0.5}$, so that an intrinsic dispersion of $\approx 0.4$ mag is
typical for a 10$^4$ M$_\odot$ cluster. It increases to $\approx 1.26$ mag
for a 10$^3$ M$_\odot$ cluster (see also Sect.\ref{sec:Montecarlo}).

An important factor in the calculation of the dispersion is the covariance
 term. Consider the definition of the (linear) correlation coefficient of
 two quantities $\rho(x,y)$ as

\begin{equation}
\rho(x,y) = \frac{{\mathrm{cov}}(x,y)}{\sigma_x \; \sigma_y}
\end{equation}

It varies between $-1$ and $+1$ where the sign indicates the sense of the
correlation, so that when $\rho(x,y)$ is equal to 1 the quantities are
completely correlated and 0 if there is no correlation.  For one given
star, as pointed out by \cite{Buzz93}, the luminosities in different bands
are completely correlated since they are produced by the same star.  But
the situation changes when several stars, instead of an individual one, are
considered, i.e. the case of population synthesis.  This is shown in the
lower panels of Fig. \ref{fig:app1}, where the corresponding $\sigma$(V--K)
values are shown for different assumptions made about the covariance. The
assumptions made are (1) no inclusion of covariance terms (or equivalently
${\mathrm{cov}({\mathrm{V,K}})} = 0$ and so $\rho({\mathrm{V,K}})=0$), (2)
a complete correlation, $\rho({\mathrm{V,K}})=1$, and (3) the case where
the covariance is obtained from the model adding properly the contribution
of {\it each} population of synthetic stars.  The figure clearly shows that
the use of $\rho({\mathrm{V,K}})=0$ leads to an overestimate of the error
while the use of $\rho({\mathrm{V,K}})=1$ produces an underestimate. Note
that this trend is independent of the presence or otherwise of the nebular
continuum.

\begin{figure}
 \resizebox{\hsize}{!}{\includegraphics[angle=270,width=7cm]{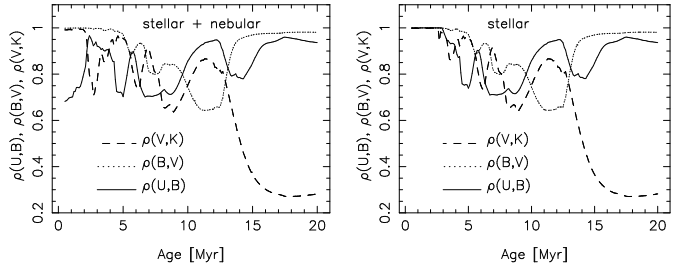}}
 \caption{Correlation coefficients. Evolution of the correlations
 coefficients for U and B, B and V, and V and K for a cluster where the
 nebular contribution is included (left) and for a cluster with the stellar
 contribution only (right).}
\label{fig:app2}
\end{figure}

\cite{LM99} argue that a complete correlation in stellar clusters can only
be guaranteed for nearby continuum bands. We have tested this assumption in
Fig. \ref{fig:app2}, where the correlation coefficient of some colour
indexes have been plotted.

The figure shows that, as expected, $\rho({\mathrm{V-K}})$ deviates from
unity, which means that the covariance term must always be taken into
account. It also shows that a similar situation happens with
$\rho({\mathrm{U-B}})$ and $\rho({\mathrm{B-V}})$. The effects become even
more severe where the nebular emission is included (especially for U--B,
since the U filter contains the Balmer jump). Therefore the assumption of
complete correlation ($\rho(x,y)=1$) produces an underestimate of the
dispersion even in close continuum bands. This result highlights the
importance of the proper inclusion of the covariance term at all
wavelengths.

\subsection{H$\beta$ equivalent width and WR ratios}

\begin{figure}
 \resizebox{\hsize}{!}{\includegraphics[angle=270,width=7cm]{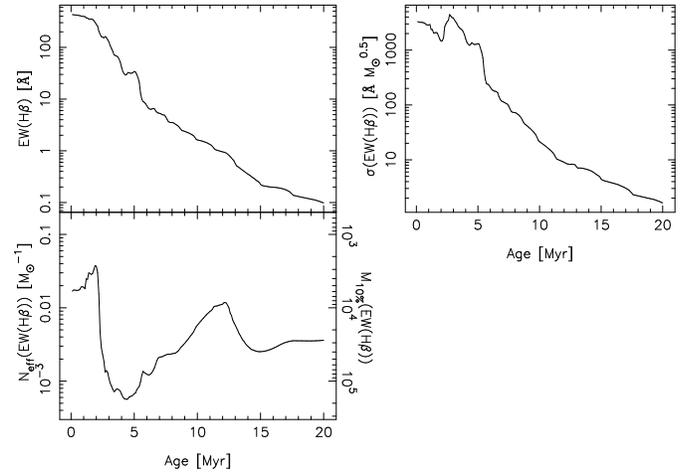}}
 \caption{Equivalent width of H$\beta$. {\it Top left:} Evolution of
 EW(H$\beta$) {\it Bottom left:} Effective number, in units of
 M$_\odot^{-1}$, for EW(H$\beta$).  Right axis as in Fig.~\ref{fig:nlyc}.
 {\it Top right:} $\sigma$(EW(H$\beta$)) in units of \AA ~M$_\odot^{0.5}$.}
 \label{fig:ewhb}
\end{figure}

Figure \ref{fig:ewhb} shows the evolution of the EW(H$\beta$), and the
associated $N_\mathrm{eff}$(EW(H$\beta$)) and $\sigma$(EW(H$\beta$)). As in
the case of colours, the scaled $\sigma$(EW(H$\beta$)) is obtained dividing
the square root of the variance by the square root of the total mass
transformed into stars.

The relative dispersion in EW(H$\beta$) is similar to the dispersion in the
continuum at such wavelengths (c.f. Fig. \ref{fig:L}) except for the very
early phases of the burst. From 0 to 2 Myr L(4861\AA) and Q(H$^+$) are
highly correlated via the influence of the nebular continuum and it gives
raise to a lower dispersion on EW(H$\beta$).

\begin{figure*}
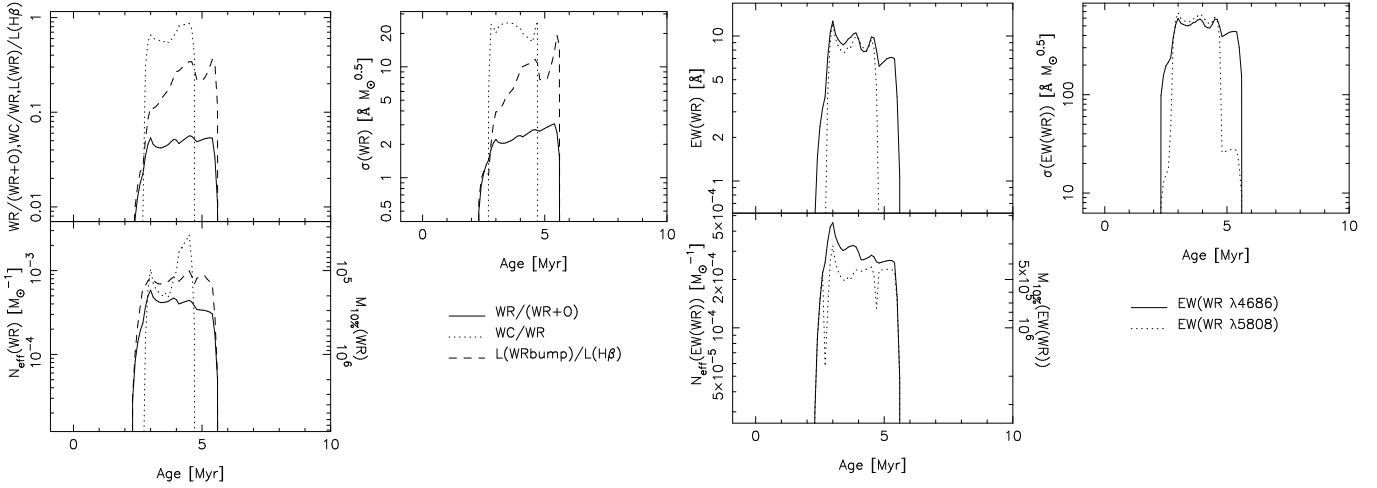

\resizebox{\hsize}{!}{\includegraphics[angle=270,width=7cm]{h2837f8a.eps}
\includegraphics[angle=270,width=7cm]{h2837f8b.eps}}
\caption{{\bf Left Panel: WR ratios} {\it Top left:} Evolution of the WR
bump over H$\beta$ luminosities ratio, WR over (WR+O) ratio and WC over WR
ratio.  {\it Bottom left:} Effective number, in units of M$_\odot^{-1}$,
for the given observables. Right axis as in Fig.~\ref{fig:nlyc}.  {\it Top
right:} $\sigma$ in units of M$_\odot^{0.5}$ (Note the normalisation
units).  {\bf Right Panel: WR equivalent widths} {\it Top left:} Evolution
of the EW(WR) for the 4686\AA ~band and 5808 \AA ~line.  {\it Bottom left:}
Effective number, in units of M$_\odot^{-1}$, for the given observables.
{\it Top right:} $\sigma$(WR bump) and $\sigma$(5808 \AA ~line) in units of
M$_\odot^{0.5}$.}
\label{fig:wrhb}
\end{figure*}

Finally the left panel of Fig. \ref{fig:wrhb} shows the evolution of the
blue WR bump at 4686\AA\ and H$\beta$ luminosity ratio, the WR over (WR+O)
and WC over WR ratios, as well as the corresponding $N_\mathrm{eff}$ and
$\sigma$ values. The right panel shows similar quantities for the
equivalent widths of the 4686~\AA\ WR band and 5808~\AA\ lines.

The figure shows that the L(WR~bump)/L(H$\beta$) ratio is more reliable
than the equivalent widths of the WR band and 5808~\AA\ line to determine
the presence of a WR population in starburst galaxies (where WR population
ratios can not be obtained directly). This is due to the fact that the WR
population itself produces ionising photons, and so it is correlated to the
H$\beta$ luminosity as well.

\section{Comparison with Monte Carlo simulations}
\label{sec:Montecarlo}

The examples in the previous section show the usefulness of the formalism
at evaluating, in a simple way, the dispersion in integrated properties,
and the insight that they provide on the effects of sampling fluctuations
in observables. The question that remains to be checked is whether the
estimates of the dispersion provided by the formalism are realist.

This issue can be solved performing Monte Carlo simulations where the
actual distribution function of a given observable is computed and its
dispersion evaluated. As an example we show here simulations for the V--K
colour index and EW(H$\beta$). This allow us to test to which extent the
formalism and the influence of the covariance terms are correct.

The simulations have been done using 1000 clusters with
$N_\mathrm{tot}$=10$^3$ stars, 500 clusters with 10$^4$ stars and 100
clusters with 10$^5$ stars. In each set we have obtained the dispersion
$\sigma_{\mathrm{clus}}^*$(V--K) and
$\sigma_{\mathrm{clus}}^*$(EW(H$\beta$)), as well as
$N_\mathrm{eff}^*$(EW(H$\beta$)).  The $\sigma_{\mathrm{clus}}^*$ values
have been divided by $N_\mathrm{tot}^{0.5}$, to obtain a normalised
$\sigma_{\mathrm{clus}}$ value for each set.
$N_\mathrm{eff}^*$(EW(H$\beta$)) has been divided by $N_\mathrm{tot}$ to
obtain a scaled $N_\mathrm{eff}$(EW(H$\beta$)) value which can be compared
with the analytical results\footnote{Note that in the simulations a fixed
total mass for the cluster cannot be achieved, since it is a random
variable. Hence, a renormalisation to $N_\mathrm{eff}$= 1 star is used
rather than to $M_\mathrm{tot}$= 1 M$_\odot$ (i.e. we have used a factor
0.17 = $N_\mathrm{tot}$ /$\lacute M_{\mathrm tot} \racute$ for the
conversion).  }.

\begin{figure*}
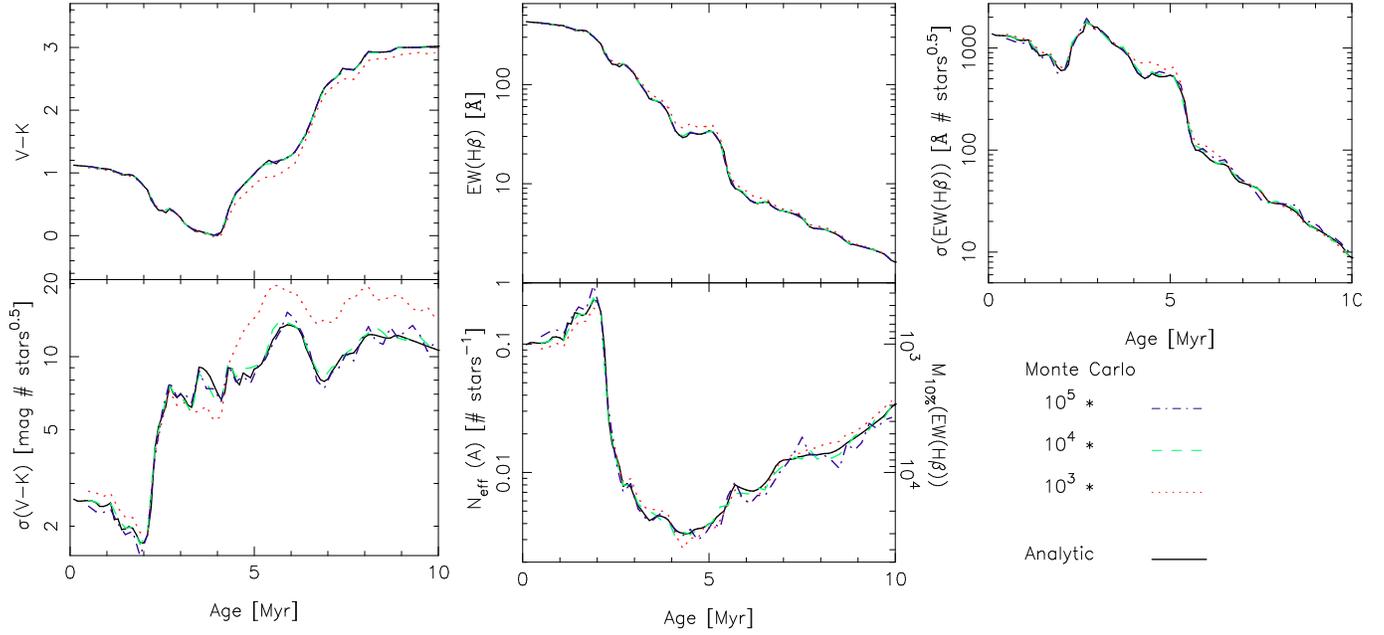

 \resizebox{\hsize}{!}{\includegraphics[angle=270,width=4.6cm]{h2837f9a.eps}
\includegraphics[angle=270,width=9.4cm]{h2837f9b.eps}}
\caption{Comparison with Monte Carlo simulations: {\bf Left Panel: V--K
color index} {\it Top left:} Evolution of the V--K color index for the
analytical and Monte Carlo simulations.  {\it Bottom left:} Comparison of
the normalised value of $\sigma$(V--K) for the analytical and Monte Carlo
simulations. {\bf Middle and Right Panels: EW(H$\beta$)}. {\it Top middle:}
Evolution of EW(H$\beta$) for the analytical and Monte Carlo simulations.
{\it Bottom middle:} Normalised $N_\mathrm{eff}$(EW(H$\beta$)) for the
analytical and Monte Carlo simulations.  {\it Top Right:} Comparison of the
normalised value of $\sigma$(EW(H$\beta$)) for the analytical and Monte
Carlo simulations.}
\label{fig:Mon}
\end{figure*}

The top left panel of Fig. \ref{fig:Mon} shows the comparison of the mean
V--K colour index for the different simulations and the analytical
solution.  The values are clearly consistent with each other, and show not
only that the analytical formalism is a very good approximation, but also
that the formalism can be used in place of Monte Carlo simulations.

There are, however, some deviations in the 10$^3$ stars cluster
simulations. As shown in \cite{CMH94} the V--K colours at the ages where
the major discrepancies appear are mostly dominated by the presence of
RSGs.  The differences between the analytical estimation of the dispersion
and the real one (as obtained from Monte Carlo runs)is perhaps due to the
low number of stars responsible of the colours. For such a small number of
stars the sampling is too poor to be meaningful.  In these low mass
clusters the analytical dispersion provides only a first order
approximation to the real one, that can only be obtained from Monte Carlo
simulations.  The lower left panel shows the comparison of the normalised
values of $\sigma_{\mathrm{clus}}$(V--K) with the analytical value.  It
shows that the analytical dispersion and the Monte Carlo results are very
similar.

We also show in Fig.~\ref{fig:Mon} the 90\% confidence levels of the
different simulations. It is important to highlight that these confidence
level bands are not {\it symmetric} with respect to the mean value.

\begin{figure*}
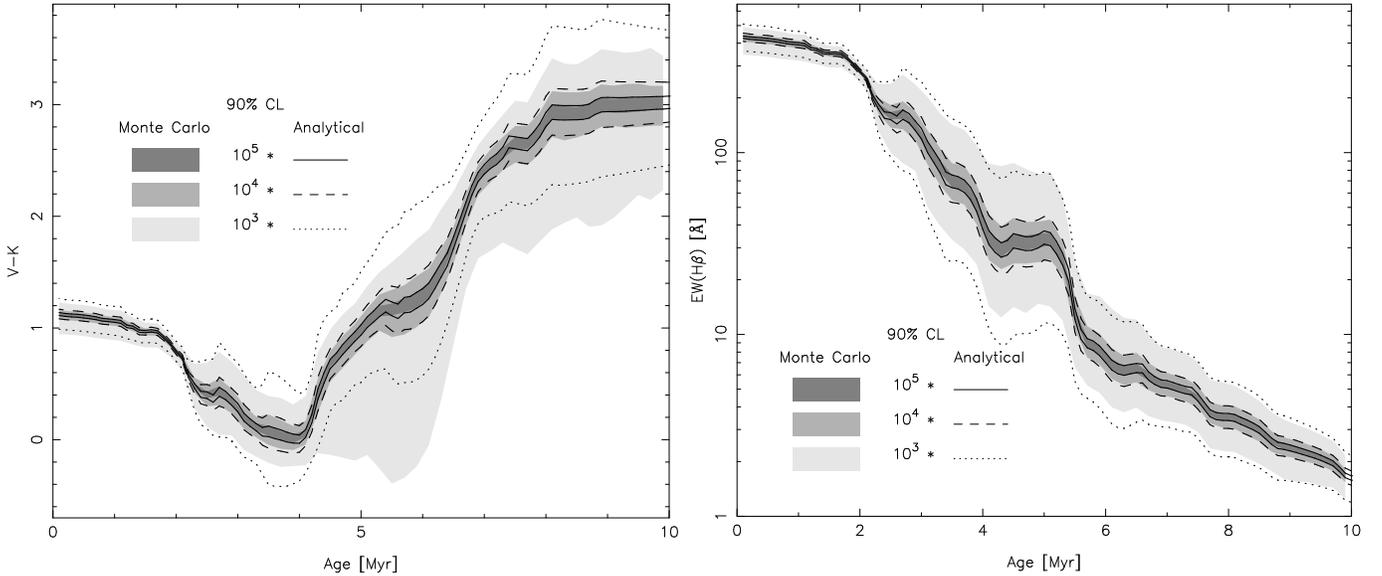

 \resizebox{\hsize}{!}{\includegraphics[angle=270,width=7cm]{h2837f10.eps}
\includegraphics[angle=270,width=7cm]{h2837f11.eps}}
\caption{Comparison of analytical confidence levels with Monte Carlo
simulations. {\it Left Panel}: V--K colour index.  {\it Right Panel}:
EW(H$\beta$)}.
\label{fig:MonCL}
\end{figure*}

It is also possible to obtain confidence level bands analytically, as shown
in Fig. \ref{fig:MonCL}.  The analytical confidence levels have been
obtained using the approximations found by \citet{Geh86} for Poisson
statistics. We have used the corresponding denormalised $N_\mathrm{eff}$,
(i.e. $N_\mathrm{eff}^*$) for the EW(H$\beta$) and the
$\mathrm{L}_\mathrm{V}/\mathrm{L}_\mathrm{K}$ observables as the Poisson
parameter.  The resulting confidence levels have been multiplied by
$A/N_\mathrm{eff}^*$ with $A$ corresponding to EW(H$\beta$) or
$\mathrm{L}_\mathrm{V}/\mathrm{L}_\mathrm{K}$. In the case of V--K, the
resulting confidence level of $\mathrm{L}_\mathrm{V}/\mathrm{L}_\mathrm{K}$
has been transformed into the corresponding V--K value.

The analytical estimation of confidence levels for clusters with masses
larger than 10$^3$ stars provides a very good approximation to the one
obtained in the Monte Carlo simulations, showing, a posteriori, that the
actual probability distribution function is well approximated by a Poisson
distribution.

\section{Discussion and conclusions}
\label{sec:dis}

In this paper we have reviewed the effects of sampling fluctuations in
evolutionary synthesis codes. We have presented an approximate, and yet
accurate, analytical formalism which allows the estimation of Poisson
dispersions for any observable synthetised by population codes.  It is
important to take into account that these dispersions vary from observable
to observable and cannot be extrapolated. We provide some simple cases
(Appendix A) for a variety of observables. The comparison of these
analytical dispersions with proper Monte Carlo simulations shows that the
formalism is valid, and hence bypasses the need to perform Monte Carlo
simulations to evaluate the dispersion in any given observable. The
formalism is simple enough that it can (and must) be implemented in all the
synthesis codes to assess the predicted dispersions in the computed
observables.  We also stress the importance of evaluating the covariance
terms when estimating the Poissonian dispersion.

As an illustrative example of the formalism, we have implemented it for the
case of young starbursting clusters. We find that in general the UV
observables are the best defined ones (i.e. they have the lowest Poissonian
dispersions). The less reliable continuum range corresponds to the hard
ionising flux.  In particular, a relative dispersion smaller than 10\% for
the hard ionising flux can only be achieved for clusters more massive than
5 10$^4$ M$_\odot$. The situation becomes even worse when WR stars appear
in the cluster. In that case a mass of about 10$^6$ M$_\odot$ is needed to
ensure an uncertainty of about 10\%.  The implications of such effects on
the use of evolutionary synthesis models as input for photoionisation ones
remain to be evaluated.

In the case of WR stars, we have shown that the most reliable observable
for the determination of the WR population is the ratio between the WR bump
at 4686~\AA\ and the H$\beta$ luminosities. We have also shown that the
presence of WR stars is a quite poorly-defined property of star forming
regions, due to small number statistics. In particular, it is possible that
the general failure of synthesis models to predict the WR population in our
Galaxy could be solved taking into account the dispersion in real clusters
due to Poissonian fluctuations in addition to the dispersion introduced by
the star formation history.

 We have also shown that the dispersions and confidence levels derived
analytically are in good agreement with the dispersions obtained from Monte
Carlo simulations for clusters containing more than 10$^3$ stars, i.e. more
than $5 \times 10^3$ M$_\odot$ transformed into stars in the mass range 2
-- 120 M$_\odot$ following a Salpeter IMF. The confidence levels and
dispersions obtained can be used as a first order evaluation of the number
of Monte Carlo simulations needed to obtain reliable results, or used
directly in the comparison with observed data.

\begin{acknowledgements}
We acknowledge the pioneering work by Alberto Buzzoni which was the
inspiration for this study.  This research has made use of the LEVEL5
utility at NASA/IPAC Extragalactic Database (NED) which is operated by the
Jet Propulsion Laboratory, California Institute of Technology, under
contract with the National Aeronautics and Space Administration.  MC has
been supported by an ESA postdoctoral fellowship.  JMMH has been partially
supported by Spanish CICYT grant ESP95-0389-C02-02.
\end{acknowledgements}

\appendix
\section{Observables, error propagation and the effective number
of stars $N_\mathrm{eff}$}
\label{ap:err}

To illustrate the use of the \Neff~ formalism for a variety of
astrophysical observables, we give here a few examples to help the reader.
For a complete review on error propagation, see \citet{KS77} or a
simplified version by \citet{Leo92} and its adaptation on the {\it NED}
server at {\tt http://nedwww.ipac.caltech.edu/level5}.

Let us assume a random variable $u$ which is a function of two random
variables $u=f(x,y)$. Let us also assume that $\sigma_x^2$ and $\sigma_y^2$
are the variances of these variables.  The variance of the random variable
$u$, $\sigma_u^2$, can be approximated as

\begin{equation}
\sigma_u^2 \simeq \left(\frac{\partial f}{\partial x}\right)^2 \sigma_x^2 +
\left(\frac{\partial f}{\partial y}\right)^2 \sigma_y^2 + 2
{\mathrm{cov}}(x,y)\frac{\partial f}{\partial x}\frac{\partial f}{\partial
y}
\label{eq:errorprop}
\end{equation}

\noindent where ${\mathrm{cov}}(x,y)$ is the covariance. This equation is
{\it exact} if the function $f$ is {\it linear} in the $x$ and $y$
variables.  The covariance can be related to the {\it linear correlation
coefficient}, $\rho(x,y)$, defined as

\begin{equation}
\rho(x,y) = \frac{{\mathrm{cov}}(x,y)}{\sigma_x \sigma_y}
\label{eq:cov}
\end{equation}

The correlation coefficient varies between $-1$ and $+1$ where the sign
indicates the sense of the correlation.

We can give a more general definition of the effective number of stars
$N_\mathrm{eff}$, as introduced by \citet{Buzz89}, to any observable. As it
was pointed out in Sect. 2, we can define a $N_\mathrm{eff}(A)$ of any
synthesized quantity $A$ which is the sum of the contributions from
individual stars (or populations), $A = \sum \omega_i a_i$, and thus it has
a variance $\mathrm{var}(A) = \sum \mathrm{var}(\omega_i) a_i^2$
(c.f. Eq. \ref{eq:cte}). The relative error is then

\begin{equation}
\frac{\mathrm{var}(A)^{1/2}}{A}= \frac{\left(\sum \omega_i
a_i^2\right)^{1/2}}{\sum \omega_i a_i} = \frac{1}{\sqrt{N_\mathrm{eff}(A)}}
\label{eq:appneff}
\end{equation}

Note that if $A$ is normalised to the total mass, $N_\mathrm{eff}(A)$ is
also normalised. So $N_\mathrm{eff}(A)$ gives an estimate of the relative
error (and hence, of the absolute error) for any possible value of the mass
transformed into stars.

\subsection{Error in the multiplication by a constant}

Let us assume an observable random variable $u(x)=a \, x$ where $a$ is a
constant.  with Eq. \ref{eq:errorprop} one obtains

\begin{equation}
\sigma_u^2 = a^2 \, \sigma_x^2
\label{eq:cte}
\end{equation}

\noindent and the relative error

\begin{equation}
\frac{\sigma_u}{u} =
\frac{\sigma_x}{x}=\frac{1}{\sqrt{N_\mathrm{eff}(x)}}=\frac{1}{\sqrt{N_\mathrm{eff}(u)}}
\label{eq:ctesig}
\end{equation}

This is the more general case used in synthesis codes, as it is showed in
Eq. \ref{eq:lcov}.

\subsection{Error of a sum}

Let us now assume an observable random variable $u= x + y$. In this case

\begin{equation}
\sigma_u^2 = \sigma_x^2 + \sigma_y^2 + 2 {\mathrm{cov}}(x,y)
\label{eq:sum}
\end{equation}

The value of ${\mathrm{cov}}(x,y)$ depends on the observables we use, as
shown by Eq. \ref{eq:lcov}.

An interesting case comes from the computation of the luminosity in a band,
for example $L_{\mathrm B}$, obtained by the integration of the
monochromatic flux, $f(\lambda)$, over the filter response of the band,
$R(\lambda)$:

\begin{equation}
{\mathrm L}_{\mathrm B} = \int{f(\lambda) * R(\lambda) \;{\mathrm d}\lambda} \simeq
\sum_{i} f(\lambda_{i})\, R_{\lambda_{i}}\,\Delta\lambda_i
\end{equation}

\noindent where we have approximated the value of the integral by the
resulting value from the use of the trapezium rule and the individual
monochromatic fluxes, $f(\lambda_{i})$ and band transmission coefficients,
$R_{\lambda_{i}}$. The corresponding variance due {\it only} to the error
in the monochromatic luminosities, $\sigma_{f(\lambda_i)} = \sigma_i$, can
be obtained from Eq. \ref{eq:errorprop} and using Eq. \ref{eq:ctesig} and
Eq. \ref{eq:sum}:

\begin{eqnarray}
\sigma_{{\mathrm L}_{\mathrm B}}^2 & = & \sum_{i} \left[ R_{\lambda_{i}} \,
  \Delta\lambda_i \right]^2 \sigma_{i}^2 \ \nonumber \\ & & + \sum_{i < j}
  2\,{\mathrm{cov}}\left( f(\lambda_i),f(\lambda_j) \right) \,
  R_{\lambda_{i}} R_{\lambda_{j}} \, \Delta\lambda_i \Delta\lambda_j
\end{eqnarray}

Note also that the resulting variance (and corresponding $N_\mathrm{eff}$)
must be similar to the one obtained if the synthesis code computes
${\mathrm L}_{\mathrm B}$, summing up the individual ${\mathrm L}_{{\mathrm
B}_i}$ and using Eq. \ref{eq:appneff}.

\subsection{Error of a difference}

Let us assume an observable random variable $u= x - y$. The corresponding
variance is

\begin{equation}
\sigma_u^2 = \sigma_x^2 + \sigma_y^2 - 2{\mathrm{cov}}(x,y)
\label{eq:res}
\end{equation}

A typical example in synthesis codes is the computation of colour
indexes. As an example, for the U-B colour index one obtains (see also
below for the error of a logarithm):

\begin{equation}
\sigma_{\mathrm{U-B}}^2 = \sigma_{\mathrm U}^2 + \sigma_{\mathrm B}^2 -
2{\mathrm{cov(U,B)}}
\label{eq:res2}
\end{equation}

This leads to $\sigma_{\mathrm{U-B}}^2 \simeq (\sigma_{\mathrm U} -
\sigma_{\mathrm B})^2$ if the correlation coefficient is 1, as it is
assumed by \cite{Buzz93} and \cite{LM99} for colours where the
corresponding fluxes originate from the same population (but it is not the
general case for integrated colours from different stellar populations).
Note that the error in colour indices may be lower than the error in the
corresponding colours, depending on the covariance term.

\subsection{Error of a product}
Let us assume an observable random variable $u= x\, y$. In this case

\begin{equation}
\sigma_u^2 = x^2 \sigma_y^2 + y^2 \sigma_x^2 + 2\, x\, y\,
{\mathrm{cov}}(x,y)
\label{eq:multi}
\end{equation}

\noindent so that

\begin{equation}
\frac{\sigma_u^2}{u^2} = \frac{\sigma_x^2}{x^2} + \frac{\sigma_y^2}{y^2} +
2 \frac{{\mathrm{cov}}(x,y)}{x\, y}
\label{eq:multirel}
\end{equation}

We can rewrite the relative error as a function of $N_\mathrm{eff}$. When
the correlation coefficient is zero (cov($x,y$)=0) :

\begin{equation}
\frac{\sigma_u^2}{u^2} =\frac{1}{N_\mathrm{eff}(u)}\simeq
\frac{1}{N_\mathrm{eff}(x)} + \frac{1}{N_\mathrm{eff}(y)}
\label{eq:neffmulti}
\end{equation}

while if the correlation coefficient is 1, cov($x,y$)=$\sigma_x \sigma_y$ and

\begin{equation}
\frac{\sigma_u^2}{u^2} = \frac{1}{N_\mathrm{eff}(u)}\simeq
\left(\frac{1}{\sqrt{N_\mathrm{eff}(x)}} +
\frac{1}{\sqrt{N_\mathrm{eff}(y)}}\right)^2
\label{eq:neffmulticov}
\end{equation}

In the general case:

\begin{eqnarray}
\frac{\sigma_u^2}{u^2} &=& \frac{1}{N_\mathrm{eff}(u)} \nonumber \\ &
   \simeq & \frac{1}{N_\mathrm{eff}(x)} + \frac{1}{N_\mathrm{eff}(y)} + 2
   \frac{{\mathrm{cov}}(x,y)}{x\, y}
\label{eq:neffmulti2}
\end{eqnarray}

This is the case, for example, of the computation of the dispersion in the
mechanical energy or mechanical power as given by the mass-loss of the
stars in the cluster and their terminal velocities.

\subsection{Error in a ratio}

Let us assume an observable random variable $u= x/y$ where $y$ is non zero.
The resulting equation for the relative error is

\begin{equation}
\frac{\sigma_u^2}{u^2} \simeq \frac{\sigma_x^2}{x^2} + \frac{\sigma_y^2}{y^2} -
2 \frac{{\mathrm{cov}}(x,y)}{x\, y}
\label{eq:divrel}
\end{equation}

\noindent which is similar to Eq. \ref{eq:multirel} but changing the sign
which corresponds to the covariance term.

In this case the normalisation is very important. Since the normalised
quantity is $\sigma^2$, then $x^2/\sigma_x^2$, $y^2/\sigma_y^2$ and $x\,y /
\mathrm{cov}(x,y)$ have the same normalisation as $x$ or $y$, and so is
$u^2 / \sigma_u^2$ which has the same normalisation as $x$ or $y$. And
since $u$ is dimensionless $1 / \sigma_u^2$ is the normalised quantity (and
not $\sigma_u^2$ as in the general case). Note that the larger the mass of
the cluster the larger should the absolute error be, which makes no sense.
It also shows the advantage of using $N_\mathrm{eff}$ instead of $\sigma^2$
for the description of errors: $N_\mathrm{eff}$ always keeps the same
normalisation.

Ratios are used a lot in synthesis codes (equivalent widths, populations
ratios, fractional contributions, etc.) so let us illustrate some
particular cases.

\begin{enumerate}
\item {\it $y=x+z$ where ${\mathrm{cov}}(x,z)=0$}. This case is
illustrative for population ratios, say the WC/WR ratio. The resulting
relative error as a function of $x,y$ is:

\begin{equation}
\frac{\sigma_u^2}{u^2} \simeq \frac{\sigma_x^2}{x^2} +
\frac{\sigma_y^2}{y^2} - 2 \frac{\sigma_x^2}{x\, y}
\end{equation}

For the particular case of ratios of population numbers it is easy to show
that $\sigma_x^2= x = N_\mathrm{eff}(x)$ and thus

\begin{equation}
\frac{\sigma_u^2}{u^2} =\frac{1}{N_\mathrm{eff}(u)} \simeq
\left|\frac{1}{N_\mathrm{eff}(x)} - \frac{1}{N_\mathrm{eff}(y)}\right|
\end{equation}

\item {\it $y=r+z$ where cov$(r,z)=0$, but cov$(r,x) \ne 0$}: This case is
illustrative for ratios of line luminosities, say the L(WRbump)/L(H$\beta$)
ratio, where the WR has a contribution to L(H$\beta$).  The resulting
relative error is

\begin{equation}
\frac{\sigma_u^2}{u^2} \simeq \frac{\sigma_x^2}{x^2} +
\frac{\sigma_y^2}{y^2} - 2 \frac{{\mathrm{cov}}(r,x)}{x\, y}
\end{equation}

In this case, we need to know the value of ${\mathrm{cov}}(r,x)$, say the
contribution of WR stars to the H$\beta$ luminosity, in order to evaluate
the error.

\item {\it $y=r+z$ where cov$(r,z) \ne 0$, but cov$(r,x) \ne 0$}: This case
is illustrative for ratios of equivalent widths, where, for example, $x$ is
the L(H$\beta$), $r$ is the nebular continuum and $z$ the stellar one.  The
resulting relative error is

\begin{eqnarray}
\frac{\sigma_u^2}{u^2} &\simeq &\frac{\sigma_x^2}{x^2} + \frac{\sigma_r^2 +
\sigma_z + 2\, {\mathrm{cov}}(r,z)}{y^2} \nonumber \\
 & & - 2 \frac{{\mathrm{cov}}(x,r)+{\mathrm{cov}}(x,z)}{x\, y}
\end{eqnarray}

\noindent which becomes Eq. \ref{eq:divrel}.  In the study of covariance
terms, it is not necessary to worry about which are the contributions {\it
if} the covariance terms have been properly taken into account.

\end{enumerate}

\subsection{Error in a logarithm}

Last, but not least, let us assume that $u = a \log x$ + c, where $a$ and
$c$ are constants. Using Eq. \ref{eq:errorprop}:

\begin{equation}
\sigma_u^2 = a^2\,\frac{\sigma_x^2}{x^2} = \frac{a^2}{N_\mathrm{eff}(x)}
\end{equation}

The denormalised value of the effective number of, say,
$N_\mathrm{eff}({\mathrm B})$, where B is the Johnson B band, gives us the
{\it absolute error} in the B magnitude for the given mass transformed into
stars.

For the case of colours indices, say U-B, using the Eq. \ref{eq:res2}

\begin{eqnarray}
\sigma_{\mathrm{U-B}}^2 &\simeq &(2.5\times \log_{10}(e))^2 \left [
\frac{1}{N_\mathrm{eff}({\mathrm L}_{\mathrm U})} +
\frac{1}{N_\mathrm{eff}({\mathrm L}_{\mathrm B})} \right. \nonumber \\ & &
\left. - 2\frac{\mathrm{cov}({\mathrm L}_{\mathrm U},{\mathrm L}_{\mathrm
B})} { {\mathrm L}_{\mathrm U}\,{\mathrm L}_{\mathrm B}} \right]
\end{eqnarray}

Note that a $N_\mathrm{eff}$(U-V) value cannot be obtained due to the
logarithm dependence in the luminosities. This situation is general for all
the colors indices.

\bibliographystyle{apj}

\end{document}